\documentstyle[12pt,epsfig]{article}
\begin{document}
\begin{center}
\vskip 2.0 truecm
{\bf \Large   Multiplicity distributions in $e^+e^-$ annihilation into hadrons
 and the extended modified negative binomial\footnote{This 
            work was  supported in part 
            by INTAS, contract INTAS-93-3602 } }
\vskip 1.5 truecm
 {\sc   O. G. Tchikilev\footnote{ E-mail: tchikilov@mx.ihep.su} }
\vskip 0.4cm
  {\em Institute for High Energy Physics  \\
   142284, Protvino, Russia\/}
\end{center}
\vskip 1.5 truecm
\begin{abstract}
\noindent
It is shown that  simple extension of the modified negative binomial
distribution   describes
negatively charged particle multiplicity distributions
 in $e^+e^-$ annihilation, measured in the whole phase space, as well as
 the modified negative binomial. 
\end{abstract}

\newpage
\pagestyle{plain}

  It has been shown recently  that
negatively charged particle multiplicity
 distributions in $e^+e^-$ annihilation into hadrons~[1-5] and in
the lepton-nucleon scattering~[6]
are well described by the  modified negative binomial distribution (MNBD).
It has been shown also~[7] that the MNBD and its simple extension EMNBD
quite well describe charged particle multiplicity distributions in restricted
(pseudo)rapidity intervals in $e^+e^-$ annihilation into hadrons and in
$e^+p$ collisions at HERA energies. The aim of this paper is to show that
the EMNBD describes negatively charged particle multiplicity distributions
 in $e^+e^-$ annihilation, measured in the whole phase space, as well as
 the MNBD.
 
 Let us remind that the MNBD can be
  defined by the probability generating function
\begin{equation} 
           M(x) = \sum_{n}{P_{n}x^{n}} = 
   {\biggl({{1+\Delta\, (1-\varphi (x))}\over{1+r\,(1-\varphi (x))}}\biggr)}^{k},
\end{equation}
where  $P_n$ is the probability to produce ~$n$~ particles,
\begin{equation}
   \varphi (x) = 1 - p~ (1-x)
\end{equation}
 and $k$, $\Delta$, $p$ and $r$ are parameters connected to the mean
 multiplicity $<n>$ by the relation
\begin{equation}
 <n> = k~p~(r - \Delta).
\end{equation}
 It has been assumed in the toy model proposed in~[1,3,4]  that the parameter
 $k$ is the number of
 sources of particle production at some initial stage of the interaction; these
 sources develop independently of each other according to some branching
 process (characterized by the parameters $r$ and $\Delta$) and during the
 branching process intermediate neutral clusters are produced.
  The parameter $p$ is equal to the cluster decay probability into a
 charged hadron pair and  $(1-p)$ is the probability of cluster decay into
 pair of neutral hadrons. The energy independence of the parameter
 $\Delta$ (or the product $\Delta\, p$), observed in the papers~[1,3,4,5] can
 indicate that the branching process is the pure birth branching process and
 $\Delta$ can be fixed at $-1$.
  The probability generating function for the EMNBD has also
 the form (1) with $\varphi(x)$ replaced by
\begin{equation} 
    \varphi_{2}(x) = 1 -\varepsilon_{1}(1-x)-\varepsilon_{2}(1-x^2),
\end{equation}
 where the  parameters $\varepsilon_{1}$ and $\varepsilon_{2}$ can be considered
 as the  cluster decay probabilities into one and two pairs 
 of charged hadrons  respectively.
 The EMNBD transforms into the MNBD when $\varepsilon_2 = 0$, in this case
 $\varepsilon_1 = p $.
 The probabilities
 $P_{n}$ for the EMNBD can be calculated using the  formulae given in~[7].
 
  The results of the EMNBD fits to the negatively charged particle
 multiplicity distributions in $e^+e^-$ annihilaion
into hadrons~[8-19]
   are given in the table~1.
 The parameter $\Delta$ was fixed at the value $\Delta=-1$  and the
 parameter $r$ was calculated from the mean charged multiplicity $<n>$
using the relation
\begin{equation}
 r = \Delta + \frac{<n>}{k\, (\varepsilon_1 + 2 \varepsilon_2)}\, .
\end{equation}
  The integer parameter $k$
 has been tested in the interval from 1 to 9, and the parameters 
$\varepsilon_1$ and $\varepsilon_2$
 have been assumed to be nonnegative. The errors for the parameter $k$  
 were calculated using the quadratic interpolation for the $\chi^2$
 dependence on $k$ on both sides from the $\chi^2_{min}$.
   One can see from the table~1  that the quality of the EMNBD fits 
is  qood. The $\chi^2$ for the EMNBD fits are in general smaller than 
the $\chi^2$ for
the MNBD fits~[4,5], shown in the last column of the table~1.
 It is necessary to  note that the $\chi^2/NDF$ values for the fits
 should be considered just indicative, since the full covariance matrix
 is not given in the experimental publications
  and therefore the proper treatment of the correlations
 between measurements of the neighbour multiplicities is not 
 possible.
  
The  energy dependence of the 
parameters $\varepsilon_1$ and $\varepsilon_2$ is presented in the fig.~1. 
 These parameters appear to be energy independent, if one excludes first
and last energy points with $\sqrt s$ equal $3$ and $\simeq 133$~GeV. The big
values $\varepsilon_2$ 
at these energies
can be explained by statistical fluctuations. One
should note also that for these energies practically the same
$\chi^2/NDF$ values are obtained for the EMNBD fits with the parameters
$\varepsilon_1$ and $\varepsilon_2$ fixed at the  average values
$\simeq 0.65$ and $\simeq 0.16$ respectively (not shown).

The energy dependence of the parameter $k$  for the EMNBD fits is
compared in the fig.~2 with the energy dependence 
of the parameter $k$ for the MNBD fits. The $k$ for both parametrizations
rise almost
linearly  with $\log (\sqrt s) $ at energies below $\simeq 30$~GeV and seem
to approach some asymptotic value $\sim 7$ or $6$ at higher energies. 
 
 At the present level of the experimental precisions in the multiplicity
measurements
both the MNBD and EMNBD parametrizations look more or less the same, this is
explained by 
the smallness of the $\varepsilon_2$ with respect to the $\varepsilon_1$.
 The good quality of fits is expected also for the next iteration when one
adds the term $\varepsilon_3\, (1-x^3)$, responsible for the cluster decay
 into three charged hadron pairs, to the function $\varphi (x)$; indeed
the probability $\varepsilon_3$ is expected to be smaller than the
$\varepsilon_2$. These iterations remind the Pad\'{e} approximants (ratios
of the polynomials) of the increasing order, used in the calculational
mathematics for the function approximation.
 The better precision of the measurements is needed in order to
clarify whether the MNBD and EMNBD  parametrizations are simply
the successive
approximations to the genuine multiplicity
 distribution, given by Nature or no additional iterations is needed.

 In conclusion, it is shown that the EMNBD describes negatively charged
particle multiplicity distributions in $e^+e^-$ annihilation into hadrons
as well as the MNBD. The energy dependence of the parameter $k$,
assumed to be the number of particle production sources, is similar
for both parametrizations. The energy independence of the EMNBD parameters
 $\varepsilon_1$ and $\varepsilon_2$ supports the toy model proposed in~[1,3,4].
Better precision in multiplicity measurements is needed to discriminate
between the MNBD and the EMNBD. 


\subsection*{Acknowledgements}
\vskip 3mm

 I am indebted to O.~L.~Kodolova for reading the manuscript and
critical comments.


\vskip 1cm
\newpage

\newpage
\begin{center}
\Large{\underline{Figure Captions}}
\end{center}
\vskip 2truecm
\normalsize

 Fig.1 The energy dependence of the parameters $\varepsilon_1$ and
$\varepsilon_2$
obtained from the EMNBD fits 
  to the negative charged particle multiplicity distributions in the
 $e^{+}e^{-}$
  annihilation. 
\vskip 0.5truecm

 Fig.2 The energy dependence of the parameter $k$ for the EMNBD fits
 compared to the similar dependence for the MNBD fits.
\vskip 0.5cm
\normalsize
\newpage
\newcommand{\qq}{\mbox{$Q^{2}$}}
\newcommand{\en}{\mbox{$\sqrt s$}}
\newcommand{\np}{\mbox{$n_p$}}
\newcommand{\pnch}{\mbox{$P(n_{ch})$}}
\newcommand{\nch}{\mbox{$n_{ch}$}}
\newcommand{\avm}{\mbox{$<n_{-}>$}}
\newcommand{\avn}{\mbox{$<n_{-}>$(fit)}}
\newcommand{\axi}{\mbox{$\chi^2$/NDF}}
\newcommand{\dda}{\mbox{$\times 10^{-1}$}}
\newcommand{\ddb}{\mbox{$\times 10^{-2}$}}
\newcommand{\ddc}{\mbox{$\times 10^{-3}$}}
\newcommand{\ddd}{\mbox{$\times 10^{-4}$}}
\newcommand{\dde}{\mbox{$\times 10^{-5}$}}
\newcommand{\ddf}{\mbox{$\times 10^{-6}$}}

\renewcommand{\arraystretch}{1.1}
\begin{table}[bth]
\caption{Results of the EMNBD fits to the negatively charged particle 
multiplicity distributions.   The last column gives the  
$\chi^2$ values for the MNBD fits
 taken from~[4,5].}     
\begin{center}
\begin{tabular}{|c|c|c|c|c|c|c|}
\hline
Experiment&$\sqrt s$ (GeV)&$k$&$\varepsilon_1$&$\varepsilon_2$ 
&\axi&$\chi^{2}$(MNBD)\\
\hline\hline
  & 3.0 & 1$\pm 0.67$ &0.462$\pm 0.049$ & 0.510$\pm 0.031$ & 5.6/2
                                                   &3.3\\
  & 4.0&2$\pm 0.47$ &0.612$\pm 0.029$ & 0.211$\pm 0.029$ &
 3.1/3 &7.6 \\ 
 MARKII[8]& 4.8&2$\pm 0.42$ &0.622$\pm 0.019$ &0.281$\pm 0.018$&
 8.2/3   &14\\ 
 & 7.4 &3$^{+3.8}_{-1.4}$& 0.544$\pm 0.074$ &0.159$\pm 0.069$ &5.1/4 
 &5.2  \\ \cline{1-7}
 &9.36&3$^{+0.86}_{-0.15}$&0.559$\pm 0.036$ &0.339$\pm 0.033$&2.6/7 &5.1 \\
 ARGUS[9]&10.58&4$^{+2.76}_{-0.94}$&0.596$\pm 0.119$ &0.308$\pm 0.102$ &
 1.3/7 &1.6\\ \cline{1-7}
 & 14.0&4$^{+0.42}_{-0.51}$&0.664$\pm 0.024$&0.165$\pm 0.021$&5.8/11&12 \\
 & 22.0 & 5$^{+0.58}_{-0.65}$ &0.655$\pm 0.029$  &0.141$\pm 0.026$ &
 5.2/12 & 8.2 \\
 TASSO[10]&34.8&6$^{+0.34}_{-1.01}$ &0.653$\pm 0.018$&0.111$\pm0.018$&
 7.8/16 &16\\
& 43.6 &8$^{+2.13}_{-1.05}$&0.567$\pm 0.030$ &0.086$\pm 0.034$&12.7/17 & 13 \\
\hline
 HRS[11]&29& 6$^{+0.90}_{-0.93}$&0.643$\pm 0.025$&0.141$\pm 0.23$&
 6.5/12&7.8\\ \hline
 &50&6$^{+1.21}_{-0.23}$&0.665$\pm 0.079$&0.184$\pm 0.077$&2.1/17 &2.3 \\
 &52&7$^{+0.69}_{-0.38}$&0.724$\pm 0.143$ &0.041$\pm 0.181$&6.1/17&6.1\\ 
 &55&6$^{+1.10}_{-0.24}$&0.680$\pm 0.079$ &0.170$\pm 0.081$&3.8/17&4.6\\
 &56&7$^{+1.54}_{-0.92}$&0.647$\pm 0.061$ &0.136$\pm 0.063$&11.4/17&12\\ 
 AMY[12]&57 &6$^{+1.10}_{-0.86}$&0.670$\pm 0.077$ &0.181$\pm 0.079$&
6.6/17&7.5\\ 
 &60&7$^{+1.84}_{-1.20}$&0.653$\pm 0.092$ &0.120$\pm 0.101$&6.1/18& 6.4\\
&60.8&7$^{+0.87}_{-0.81}$&0.648$\pm 0.053$ &0.167$\pm 0.053$&15.8/18& 16\\
&61.4&6$^{+0.88}_{-0.20}$&0.691$\pm 0.073$ &0.171$\pm 0.073$&9.1/18& 10\\ 
\hline
ALEPH[16]&91.2&8$^{+1.16}_{-1.46}$&0.657$\pm 0.159$&0.039$\pm 0.225$& 10.9/21
 & 11 \\ \hline
ALEPH[17]&91.2&7$^{+1.13}_{-2.00}$&0.736$\pm 0.356$&0.035$\pm 0.444$& 3.8/24
 & 3.8 \\ \hline
DELPHI[13]&91.2&7$^{+0.55}_{-1.74}$&0.754$\pm 0.203$&0.334$\pm 0.252$&18.6/19&
 30 \\ \hline
DELPHI[14]&91.2&6$\pm 0.23$&0.639$\pm 0.022$& 0.228$\pm 0.080$ & 94.2/23& 
 \\ \hline
 L3[18] & 91.2& 7$^{+0.75}_{-0.29}$ &0.696$\pm 0.107$ & 0.074$\pm 0.119$& 
14.0/21 & 14 \\ \hline
 OPAL[15]& 91.2&7$^{+0.58}_{-0.18}$&0.678$\pm 0.065$ & 0.101$\pm 0.066$& 4.2/23
 & 5 \\ \hline
 OPAL[19]&133&5$^{+1.75}_{-0.35}$ &0.433$\pm 0.150$ & 0.493$\pm 0.113$& 3.7/21
 & 5.3 \\
\hline
\end{tabular}
\renewcommand{\arraystretch}{1.0}
\end{center}
\end{table}
\clearpage
\begin{figure}[th]
\begin{center}
\epsfig{file=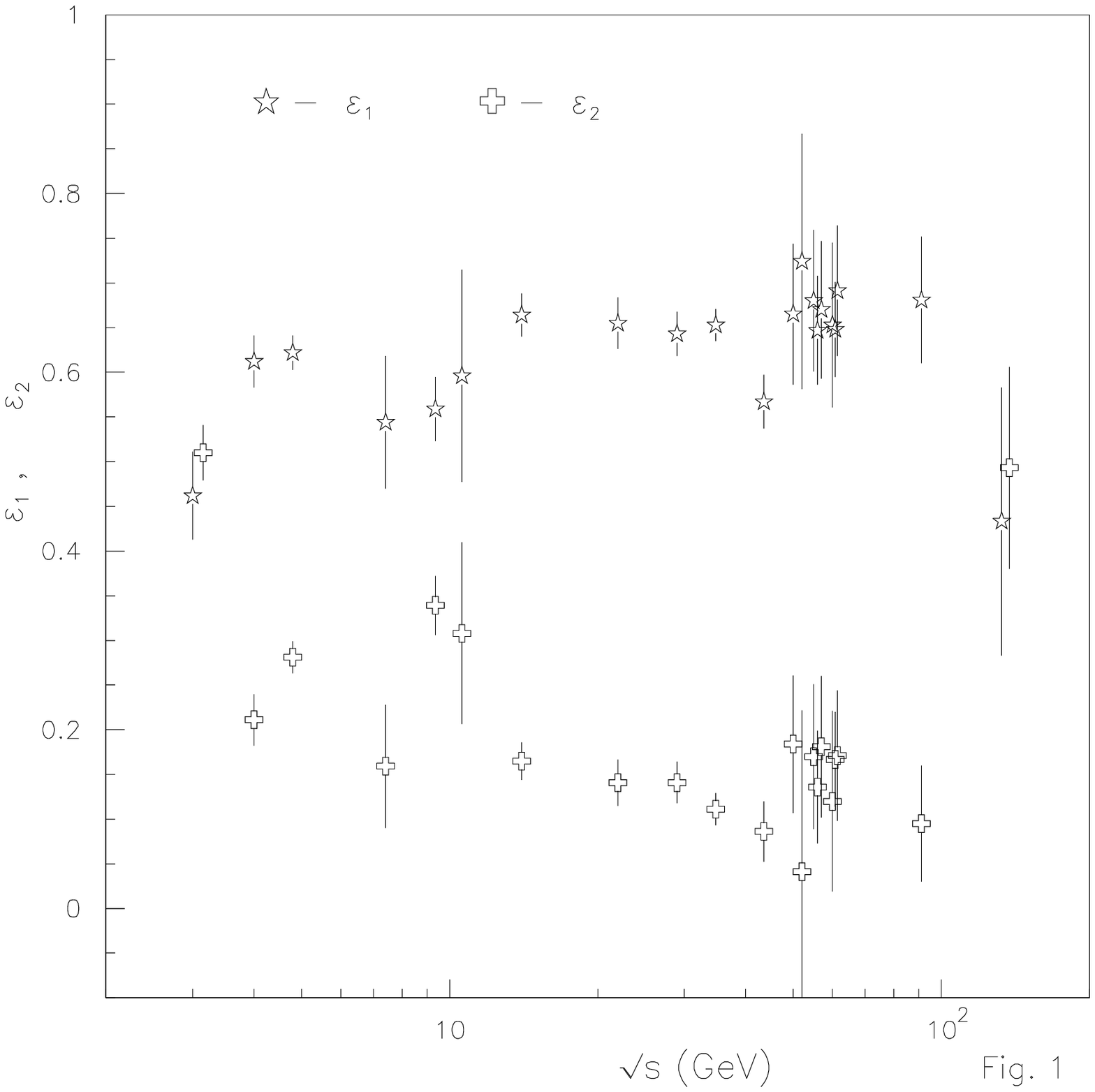,bbllx=30pt,bblly=120pt,bburx=590pt,bbury=650pt,%
width=18cm,clip=}
\end{center}
\end{figure}
\newpage
\begin{figure}[th]
\begin{center}
\epsfig{file=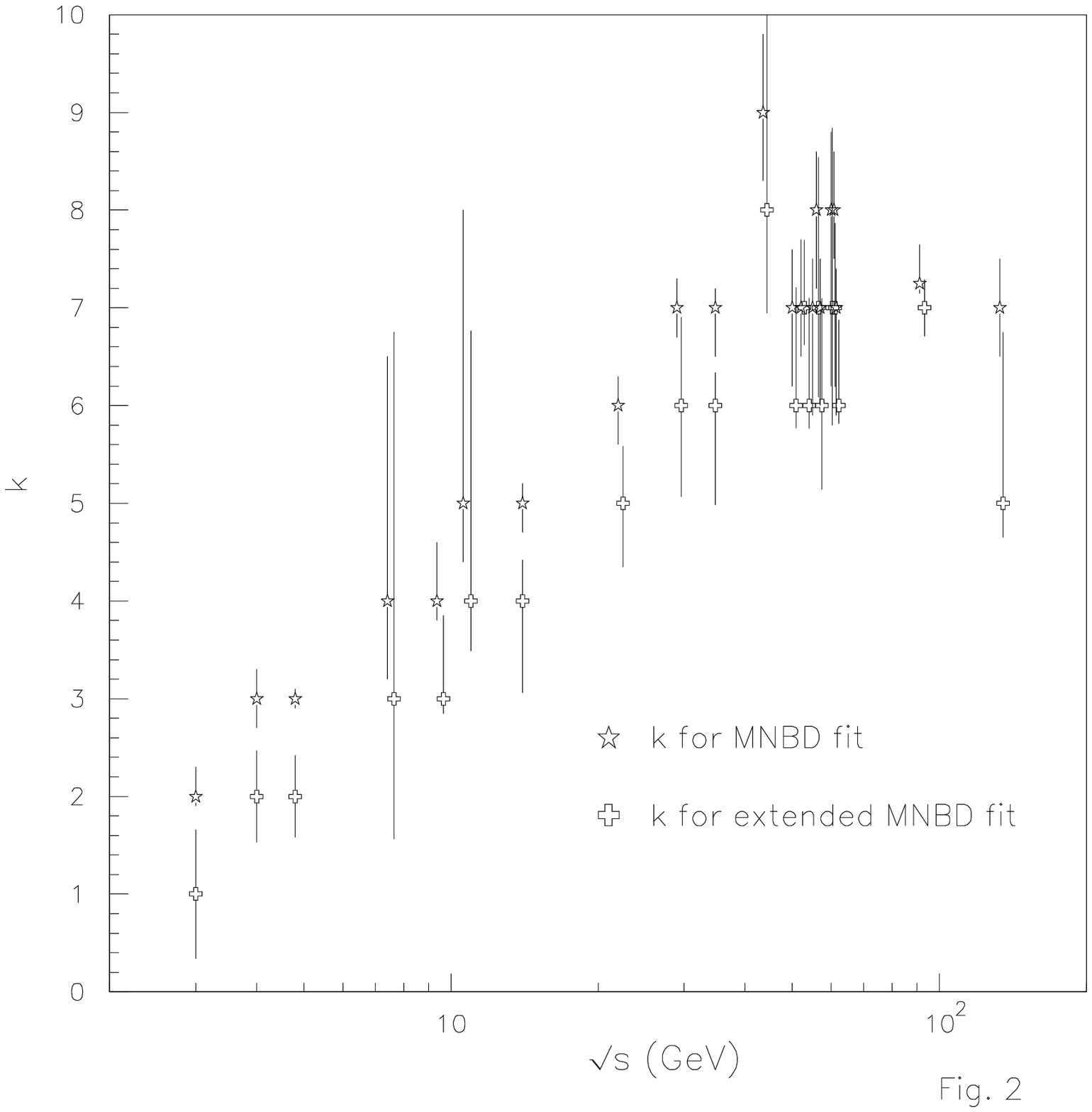,bbllx=30pt,bblly=120pt,bburx=590pt,bbury=650pt,%
width=18cm,clip=}
\end{center}
\end{figure}
\end{document}